\documentclass[sigconf]{acmart}
\usepackage{lipsum}
\usepackage{caption}
\usepackage{subcaption}
\usepackage{hyperref}

\usepackage{graphicx}
\usepackage{enumitem}

\hypersetup{
    colorlinks=true,      
    linkcolor=blue,       
    citecolor=blue,       
    urlcolor=blue         
}

\AtBeginDocument{%
  \providecommand\BibTeX{{%
    \normalfont B\kern-0.5em{\scshape i\kern-0.25em b}\kern-0.8em\TeX}}}

\copyrightyear{2023}
\acmYear{2023}
\setcopyright{rightsretained}
\acmConference[EAAMO '23]{Equity and Access in Algorithms, Mechanisms, and Optimization}{October 30-November 1, 2023}{Boston, MA, USA}
\acmBooktitle{Equity and Access in Algorithms, Mechanisms, and Optimization (EAAMO '23), October 30-November 1, 2023, Boston, MA, USA}
\acmDOI{10.1145/3617694.3623255}
\acmISBN{979-8-4007-0381-2/23/11}

\acmConference[EAAMO '23]{Equity and Access in Algorithms, Mechanisms, and Optimization}{October 30--November 1,
  2023}{Boston, MA}
\acmPrice{15.00}
\acmISBN{978-1-4503-XXXX-X/18/06}

\begin{document}

\title{Should they? Mobile Biometrics and Technopolicy meet Queer Community Considerations}

\author{Anaelia Ovalle}
\authornote{All authors contributed equally to this research.}
\affiliation{%
 \institution{they/them}
  \state{University of California, Los Angeles}
  \country{} 
 }

\author{Davi Liang}
\authornotemark[1]
\affiliation{%
  \institution{they/them}
  \state{New York University}
  \country{}}

\author{Alicia Boyd}
\authornotemark[1]
\affiliation{%
  \institution{she/her}
  \state{New York University}
  \country{}}

\renewcommand{\shortauthors}{Ovalle, Liang and Boyd}

\begin{abstract}
Smartphones are integral to our daily lives and activities, providing us with basic functions like texting and phone calls to more complex motion-based functionalities like navigation, mobile gaming, and fitness-tracking. To facilitate these functionalities, smartphones rely on integrated sensors like accelerometers and gyroscopes. These sensors provide personalized measurements that, in turn, contribute to tasks such as analyzing biometric data for mobile health purposes. In addition to benefiting smartphone users, biometric data holds significant value for researchers engaged in biometric identification research. Nonetheless, utilizing this user data for biometric identification tasks, such as gait and gender recognition, raises serious privacy, normative, and ethical concerns, particularly within the queer community. Concerns of algorithmic bias and algorithmically-driven dysphoria surface from a historical backdrop of marginalization, surveillance, harassment, discrimination, and violence against the queer community. In this position paper, we contribute to the timely discourse on safeguarding human rights within AI-driven systems by providing a sense of challenges, tensions, and opportunities for new data protections and biometric collection practices in a way that grapples with the sociotechnical realities of the queer community.

\end{abstract}




\maketitle

\section{Introduction}
\begin{quote}
``...Your scientists were so preoccupied with whether or not they could; They didn't stop to think if they should." -Dr. Ian Malcolm, \textit{Jurassic Park (1993)}
\end{quote}


Smartphones offer a wealth of functionalities that can aid everyday life. These compact yet powerful devices are versatile tools, offering digital connectivity, entertainment, and even health monitoring. Embedded in the majority of smartphones, sensors such as accelerometers and gyroscopes offer valuable signals for monitoring and providing feedback for various daily activities, including physical activity and sleep ~\citep{kroger2019privacy}. Paired with cloud technology, collecting biometric data like heart rate and daily moving patterns can be done in a systematic and comprehensive fashion. Consequently, smartphones can leverage this big data to deliver personalized user experiences at scale. This, in turn, highlights their value in enhancing daily life and subsequently, the desire for smartphone users to carry them everywhere.





Biometric data is also incredibly valuable to research involving user authentication \cite{derawi2010unobtrusive} and human activity recognition more broadly \cite{bao2004activity}. Accelerometers can measure movements and vibrations of an object, contributing to an ability to gather knowledge on a person's steps and gait. However, biometric data has also been employed in the task of predicting gender, with some works inferring this based on a smartphone's placement on a person's body ~\citep{jain2017human, meena2020gender, sabir2019gait, weiss2019smartphone}. Giving credence to a task definition intrinsically based on hegemonic definitions of gender has \textit{harmful impacts} to communities that exist outside of these presuppositions. The resulting modeling, therefore further marginalizes communities that do not reflect the normative assumptions codified within the machine learning (ML) and broader artificial intelligence (AI) pipeline. In this work, our paper centers the queer\footnote{Queer historically reflects a term to describe non-normative sexual and gender identity, and their intersections~\cite{keyes2018misgendering}.} community in examining mobile biometric design and problem formulation that, while well-intended, risks propagating social harms if sociotechnical research assumptions remain unscrutinized. 




Dr. Malcolm's famous quote anchors this position paper to the broad ethical question of whether we should continue to use collected data from our smartphones without making a concerted effort to pause and think about the societal harms and unintended consequences it might have on historically marginalized communities such as the queer community. How will having access to this biometric data possibly harm, perpetuate, and reinforce stigmas about queer bodies throughout our society? The technologies and data-driven processes we currently employ consume data inherently woven with societal assumptions and inductive biases ~\cite{johnson2021towards,Boyd2023reflexive}. Therefore, we reason that it behooves researchers to exercise caution by first taking a step back and evaluating possible sociotechnical repercussions to AI design choices.


\textit{Choice} is a mechanism by which individuals propagate their own views into research, design, algorithms, and policy dictating the use of these technologies. Choices in AI design and, similarly, technopolicy design do not exist in isolation to stakeholders' viewpoints and internal social constructs \cite{boyd2021quantitative, Boyd2023reflexive, ovalle2023factoring}. \textit{Who} is included in an evaluation is a function of \textit{what} a researcher or policy stakeholder understands should be included. As such, leaving unexamined the inextricable link between perceived social constructs and design risks harming communities which fall outside of these dictated boundaries of consideration. To demonstrate this, we show how limiting the definition of gender to that of a binary translates to a similarly narrow conceptual design in mobile biometrics and automatic gender prediction systems, having unique adverse impacts on gender minorities. 

Our paper critically examines the power \textit{choice} holds within the ecosystem AI resides in, consisting of both ML researchers \textbf{and} technopolicy makers that govern AI usage and design. Advancements of AI technologies have outpaced policy which adequately protect users from harm. We find this through the critical detailing of U.S. technopolicy gaps which result in limited protections for historically marginalized groups, such as the queer community, leaving them vulnerable to technological harms. Because of these gaps, we advocate for more comprehensive technopolicy that include state and federal statutes to ensure the safety, security, and privacy of all users with mobile devices that collect biometric information such as movement data and gait. 

Overall, this position paper aims to outline queer community considerations in mobile biometrics and AI for practicing researchers and technopolicy makers alike. We contribute the following:

\begin{enumerate}
    \item Detail the dangers of gender-normative assumptions in AI with respect to the queer community (\S\ref{sec:queerintro}).
    \item A discussion grounded on the dangers of smartphone biometric analysis for gender prediction (\S\ref{sec:biointro}, \S\ref{sec:harms}).
    \item A survey on existing and ongoing United States (U.S.) data privacy and gender discrimination law that is contextualized in the queer community and their corresponding needs (\S\ref{sec:law}).
    \item Explore ways to codify inclusive law and technological praxis (\S\ref{sec:foodforthought}) by adopting a reflexive approach to AI design and research processes ~\cite{miceli2021documenting, boyd2021quantitative, Boyd2023reflexive, ovalle2023factoring}.

\end{enumerate}

\textit{Positionality Statement}
All authors are people of color who center an intersectional perspective in their social and professional lives. With the exception of one author, all were formally trained as computer scientists and identified as queer, respectively. One author is formally trained in data privacy law and how technology can lead to discrimination. All authors have additional training in gender theory, critical social theories, and criminology. All authors have training in queer studies through activism and advocacy. As such, our backgrounds influence this work's posture. All authors are located in the U.S. but have diasporic links to other social contexts. Our positions arise with respect to U.S. law, though this work also has implications for AI in a global context. We write this to empower individuals across the existing tech industry and upcoming technopolicy to critically consider how we can collectively create mechanisms that better include and/or codify the inclusion of queer bodies. Therefore, this paper is positioned for AI industry practitioners, AI policymakers, and the queer community.

\section{Background} 
\label{sec:biointro}
\subsection{Mobile Biometrics}
The term \textit{biometrics} is described as the unique physical or behavioral characteristics that can be used for automatic recognition of individuals~\cite{weiss2019smartphone, bouchrika2018survey}. Biometrics can be physical (\textit{e.g.,} fingerprints, faceprints, ear shape, iris, and retina) or motion (how someone is walking or otherwise moving). Biometrics researchers have reasoned that biometrics is an unobtrusive way to protect a user's privacy on a cellular device~\cite{1415569}. Researchers have further illuminated the benefits of using biometrics on smartphones to secure and protect personal information (\textit{i.e.,} passwords and private information) stored on devices, especially if a device is stolen or lost ~\cite{alzubaidi2016authentication}. Therefore, mobile devices are commonly employed for identification and verification. However, researchers need accurate, abundant, and high quality mobile data for biometrics. 



Gait, a biometric based on human locomotion, is difficult for users to conceal or fake, thereby making it preferable to utilize in tasks like authentication \cite{fish1993clinical, celik2021gait}. Wan \textit{et al.}\cite{wan2018survey} name six properties which make gait a favorable biometric to collect. According to them, gait can be: (1) captured from far away (2) low resolution and therefore not costly to collect  (3) requiring low instrumentation (4) done without the user's cooperation (5) hard to impersonate and (6) more ``accessible''. ~\footnote{Authors did not specify what 'accessible' meant in their paper.} These locomotive patterns serve as valuable insights to several domains. In a clinical setting, gait analysis assists in diagnosing, preventing, and monitoring neurological, cardiac, and age-related disorders~\cite{muro2014gait}. In assessing gait, clinicians may use a combination of image processing, floor sensors, and sensors located on the body \cite{muro2014gait}. Gait follows a standard pattern within the clinical literature, though importantly, variables like shoe type and posture impact the ablity to measure a person's gait in a consistent manner. In computer science, gait has been studied for several applications, including authentication~\cite{wan2018survey}. Gafurov~\cite{gafurov2007survey} describes \textit{gait recognition} (or \textit{Biometric gait recognition}) as a way for a user to authenticate simply by walking. Historically, gait data solely consisted of recorded videos. However, the advent of smartphones brought with it a paradigm shift; built-in accelerometers and gyroscopes offered a faster and less resource-intensive avenue for gait data collection.

\subsection{Automatic Gender Recognition Systems}
Automatic gender recognition systems (AGR) are designed to determine an individual's gender based on various biometric features or behavioral patterns. These systems are commonly based on features including a person's gait, face, and voice. Gait-based gender prediction is described as a \textit{soft} biometric which significantly increases the performance of hard biometric systems that are strictly based on mobile features \cite{isaac2019multiview}. Since works by Mantyjarvi~\cite{1415569}, several studies have attempted to use gait-based gender recognition for tasks including but not limited to authentication and security. To do this, works build off Li \textit{et al.} ~\cite{li2008gait}, who detailed that human gait could be separated into distinct components based on human silhouettes. Yu \textit{et al.} ~\cite{yu2009study} approach gender recognition by looking at human static and dynamic silhouettes. Other works ~\cite{jain2017human, meena2020gender, sabir2019gait, weiss2019smartphone} have focused on using built-in sensors such as accelerometers for gait and AGR prediction based on the device's location on a device user's body.

However, as we detail below, several forms of biometric recognition software have demonstrated poor accuracy in identifying anyone outisde of a binary gender. Scheuerman \textit{et al.} ~\cite{scheuerman2019computers} found that commercial facial analytics services consistently performed worse at predicting gender for trans and nonbinary individuals than cisgender individuals. Other studies using AGR systems like Buolamwini and Gebru ~\cite{buolamwini2018gender} found reductionist views on gender that only distinguish between ``man'' and ``woman''. These examples are only brief examples on the ways issues in accuracy can arise due to \textit{choices} to propagate cisnormative viewpoints in design. We continue to further ground what AI-driven AGR systems mean for the queer community in the next section.





\section{Data-Driven Systems and AI: Histories of Harm to the Queer Community}
\label{sec:queerintro}

The mainstreaming of nonnormative genders and sexualities gives the impression that queer and trans lives and experiences are incorporated into the normative world~\cite{bey2021trouble, shapiro2015gender}. Yet, scholars~\cite{keyes2018misgendering,tomasev2021fairness,bey2021trouble,shapiro2015gender} have cautioned this visualization does not necessarily construe the achievement of equity throughout this cis-heteronormative social-technical world. The queer and trans population experiences several stressors due to the propagation of cissexist systems \cite{puckett2023systems}. Gender is complex and its definitions are conservative~\cite{collins2004black}, therefore gender can not be confined to binary constructs such as ``woman" and ``man"; this gender construct occultates the growing minority of those who fall inside and outside the binary norms. 
According to a Pew Research Center survey in 2022, 1.6\% of U.S. adults are transgender or non-binary. Those under 30 are more likely than older adults to be trans or nonbinary, at 5\% - and this is more likely than not to be reflective of future trends in gender expression~\cite{brown_2022}.

Similarly, algorithmic systems trained on such cis and heteronormative dimensions of social reality go on to perpetuate hegemonies through ever-present normativities. For instance, several studies~\cite{tomasev2021fairness, queer-in-ai-risk-management, ovalle2023queer} describe how cis and heteronormativity are not only propagated into AI systems but serve as foundations for mistrust - causing consequences pertaining to surveillance and communal harm to the queer community. Tomasev \textit{et al.}~\cite{tomasev2021fairness} describe how censorship in the form of erasing queer voices and amplifying heteronormative voices is possible, by censoring content resulting directly from being adjacent to the queer community. Simultaneously, there is a disproportionate online harassment~\cite{bailey2021misogynoir} and filtering out queer content for anti-queer propaganda~\cite{anderson2020queer,duguay2020queer} Such automated filtering reflects an erasure that reifies cissexist systems.

Sexual and gender identity are private components of one's identity. Tomasev \textit{et al.}~\cite{tomasev2021fairness} describe how possibilities to use AI risk infiltrating otherwise safe spaces in the name of facial recognition and overall surveillance for safety. Indeed, ample scholarship speaks to the ways such tools can further marginalize the queer community \cite{queerinai2023queer, dennler2023bound}. For instance, physiognomic and phrenologic applications such as computer vision to (falsely) infer gender and sexuality which is antithetical to queerness and queer bodies~\cite{stark2021physiognomic}. Surveillance in these areas that collide with poor privacy measures leads to (1) forced outing \cite{ovalle2023queer}  (2) applications upon ``determining someone's gender" (\textit{e.g.,} such as recommendation systems providing binary gender-based clothing) may lead to algorithmically-induced gender dysphoria, as well as (3) queer erasure through the privileging of cis-bodies, detection of bodily parts, and propagation of physiognomic thinking \cite{dennler2023bound}.

Hamidi \textit{et al.}~\cite{hamidi2018gender} found that transgender individuals have overwhelmingly negative attitudes to automated gender recognition systems due to their impact on misgendering individuals. Most AGR systems had performance issues when tested against transgender individuals, and did not classify nonbinary and queer people outside of cisnormative labels~\cite{scheuerman2019computers}.  These AGR systems cause dysphoria by failing to recognize the chosen gender of transgender individuals accurately and do not acknowledge nonbinary individuals, creating labeling issues that queer users might not desire, with discrimination followed by making automated decisions based on incorrect gender predictions. Such labeling has been theorized to bind concepts and labels like ``makeup" to gender identity, and lead to bias by third parties who use gendered appearances. Other researchers like Mahalingam and Ricanek \cite{mahalingam2013eye} have used video recognition algorithms to try and determine whether a person has gone through hormone replacement therapy. These efforts have been criticized for compiling their data from transgender users on YouTube without notice or consent and for the potential ethical issues of creating an algorithm with the power to distinguish between cis and trans people, enabling harassment and governmental persecution~\cite{Vincent_2017,Dazed_2019}.





In the following section, we take a closer look at ways in which mobile biometric technologies have propagated these aforementioned harms through \textit{choices} in their task design.


\section{Biometrics, Choices, and Their Impact on the Queer Community}
\label{sec:harms}
Gender recognition systems, at their core, hinge on the notion of extracting gender information from inherently unobservable and often superficial characteristics \cite{tomasev2021fairness}. This problematic foundation perpetuates an understanding of gender as a static, physiognomic concept, thus failing to consider gender as a social construct. As a result, the diversity of gender identities and expressions do not exist beyond the confines of a binary. These systems, built upon the notion of distinct features reflecting a ``man' and a ``woman'', tend to disregard the complex and nuanced reality of gender. Keyes~\cite{keyes2018misgendering} urges researchers to consider how they assess and conceptualize the term gender critically. Too often, they argue, gender is treated explicitly or implicitly as a “binary, immutable and physiologically discernible concept'' in research. Yet, restricting gender to a fixed, static context erases transgender and non-binary people, cascading throughout the design and research process. Albert and Delano~\cite{albert2021whole}, scholars in bio-impedance technology\footnote{Bio-impedance Technology is referring to a method for estimating body composition}, an intersecting field with biometrics, similarly discuss how it excludes non-binary and trans people. Their discussion is relevant to mobile devices. The researchers point out the field's assumptions in how sex and gender are straightforward. As we delve further into the subsequent sections, our critical examination reveals how these normative assumptions not only persist but also propagate throughout the various tasks and applications of these systems, ultimately perpetuating harmful biases and inaccuracies that have far-reaching consequences. These sections all converge towards questioning whether researchers should be making such design choices followed by critically discussing their links to technopolicy.





\subsection{Reconstructing Personal Body Data from Mobile Sensors}

Biometrics studies have shown that accelerometer data can be used to reconstruct "sensitive personal data" from its users~\cite{kroger2019privacy, zuboff2019surveillance}. While accelerometers are predominantly used in mobile devices for various applications like cameras and microphones, inferences about an individual's identity, demographic, personality, and activities can intrude on their privacy. Accelerometers can be combined with other biometric data-gathering devices and data sources to form multimodal signals which more accurately predict human activity than one single device~\cite{aguileta2019multi}. Kroger~\textit{et al.} \citep{kroger2019privacy} describe accelerometers as ``cheap, low in power consumption, and often invisibly embedded into consumer devices. Thus, they represent a perfect surveillance tool as long as their data streams are not properly monitored and protected from potentially untrusted parties including  service providers and app developers."~\citep{kroger2019privacy}. At the same time, safeguarding privacy is essential to uphold individual autonomy, particularly for those whose gender identity could expose them to persecution if revealed. This practice brings to light poorly regulated access to information about users' physical biometrics and identity that can be used to harm them.

\subsection{Determining Gender from Clothing}
Some AGR research reinforces the idea that an individual's clothing choices can be used to predict binary gender identity. They may contend that the types of clothing individuals wear, if they contain pockets ~\citep{nachtigall2018making, myers2014picking} or certain footwear ~\citep{Cosgrove_2019}, could offer significant insights into their gender. As such,  works like Friere-Obregon \textit{et al.}\cite{freire2014automatic} and Guan \textit{et al.}\cite{guan2012robust} aim to derive user information on based on clothing. However, it is essential to approach this research with caution, as it risks oversimplifying a deeply personal and multifaceted aspect of an individual's identity. Predicting gender solely from clothing data can reinforce stereotypes and undermine the lived experiences of those within the queer community and others whose gender expression may not conform to societal norms. 



Remarkably, multiple studies highlight the inherent variability in clothing-based gender prediction. Yet, rather than avoiding this challenge altogether, some researchers opt to address it by incorporating an additional task to enhance the effectiveness of clothing-based AGR systems. Guan \textit{et al.} ~\cite{guan2012robust} observed that a person's gait could change with the clothing type, reducing the effectiveness of gait recognition. The researchers designed a method to address this concern through the creation of a dataset encompassing a fixed set of clothing. Ghebleh and Moghaddam ~\cite{ghebleh2018clothing} also found difficulties in capturing human gait patterns using accelerometers due to the variability in user clothing. Because of this, they employed various outlier detection techniques for ``mitigation''. \textit{Choosing} to build technologies on others that presuppose binary gender reflects the gross magnification of gender hegemonies, reinforcing the binary concept of man or woman. As we continue to detail works and their complexities with respect to gender, it becomes increasingly clear that we need a robust technopolicy framework to safeguard the well-being of the queer community.




\subsection{Determining Gender from Smartphone Location on the Body}

A research task that infers gender based on smartphone placement on user bodies is infused with harmful assumptions on gender and bodily autonomy. While such endeavors might not appear harmful in theory, the task insidiously perpetuates the notion that gender can be ``detected'', determined by passively collecting data on someone's body, and that it should be detected. Biometric AGR works like Abuhmad \textit{et al.} \cite{abuhamad2020sensor} describe gait-based user authentication methods as being ``feasible in specific applications, which requires capturing the user gait traits while moving''. Works also explore the usability of the ``user’s physical state'', without detailing how this is defined nor its implications on user privacy. As with the above example on clothing, users carry their smartphones in pockets, but not all users have pockets in every garment they wear. If a user wears a bra or a purse, they may use the garment to carry their phones \cite{ichikawa2005s}. However, it is crucial to recognize the inherent privacy risks and ethical concerns associated with this reductive approach. First, predicting gender from such data center cisnormativity and bioessentialism, thereby erasing any representation which outside a binary. Second, it does this in an incredibly invasive manner. Rather than pursuing research that could inadvertently marginalize the queer community, it is imperative that we prioritize community concerns in their development, paired with robust technopolicy that ascertains the privacy and dignity of all individuals across gender identity and expression.

\section{Existing Data Privacy Law \& Jurisprudence}
\label{sec:law}
\subsection{Background}

The sexual orientation and gender identity (SOGI) of LGBTQIA+ people and communities have been historically criminalized and persecuted ~\cite{woods2017lgbt}. Today, being publicly outed by a release of SOGI data can lead to being disowned, discriminated against in public accommodations, discriminated against in employment and social life, and even lead to physical threats \cite{FPF2022SOGI}. While more LGBTQIA+ protections have been advanced in the U.S., many states still have or are in the process of creating laws to prohibit LGBTQIA+ expression or rights or have insufficient protections against discrimination \cite{AmericanCivilLibertiesUnion_2023}. In this environment, protecting one's privacy and identity is important for preventing data privacy harms associated with their outing and identity self-determination \cite{citron2022privacy}. This section briefly examines key data privacy protections in the European Union (E.U.) and the U.S. and examines the existing gaps for protecting queer people. 

Guarantees for data privacy exist in several forms. The most prominent are data protection laws like the E.U.’s General Data Protection Regulation (GDPR) which offer a comprehensive set of regulations on companies that collect data, such as how consent must be acquired, what data can be shared, how data can be processed, and what rights consumers have to their data \cite{wolford2020gdpr}.
The GDPR identifies several categories of sensitive personal data, including racial or ethnic origin, political opinions, health data, sex life, or sexual orientation without consent, legitimate purpose, or other key exemption~\footnote{General Data Protection Law, federal law 13.709/2018, Art. 9.}. The key feature of the GDPR is the standard for requiring consent from data subjects before data processing may occur, with heightened requirements for explicit consent before sensitive personal data like biometrics can be processed~\cite{ashman2020outed}.
While gender identity or transgender status is not explicitly included within Article 9, details about transgender status like gender reassignment or sex life will fall within the protections of Article 9 of the GDPR \cite{InformationCommissionersOfficeUK}. Importantly for transgender people, the GDPR also includes a "right to be forgotten," a form of user control over personal information on the internet to remove or correct information stored online. This can be of particular value to transgender people who seek to have their gender identity updated to reflect their current identity~\cite{correia2021gender}\footnote{The right to be forgotten can provide transgender individuals with the additional data privacy power to conceal their past, offering greater self-determination over their identity and helps to mitigate risks of being outed and targeted for their transgender history than other privacy laws~\cite{iannucci2020erasing}. Some authors like Garcia-Murillo and McInnes raise issues with the right to be forgotten over the difficulties involved with prior data leaks, issues with deleting data stored by third parties not subject to right to be forgotten laws, when a data controller has a valid exemption listed in Article 17 of the GDPR, and ethical issues over information that would be useful to others or our future selves, such as health or biographical information~\cite{garcia2018cosi}. These authors instead suggest alternate solutions of changing social norms or advancing anti-discrimination legislation, but enacting those changes can be more difficult than furthering data privacy law with a right to be forgotten. Furthermore, enforcing the right to be forgotten can be difficult and time-consuming with the number of data processors that interact and distribute any individual's data.}.  

There are other data protection laws for other countries. Many are modeled on the GDPR, like Brazil’s General Data Protection Law (LGPD)~\cite{digitalguardianBreakingDown} or the California Consumer Privacy Act (CCPA)~\cite{pardau2018california}. The U.S. does not have a comprehensive data protection law on the federal level. Instead, the U.S. relies on data protection laws for specific forms of information or content (\textit{e.g.,} Health Insurance Portability and Accountability Act and the Illinois Biometric Information Privacy Act), classes of people (\textit{e.g.,} laws protecting minors like the Children's Online Privacy Protection Act), and data protection laws in individual states (\textit{e.g.,} California Consumer Privacy Act).

Aside from data protection laws, other privacy laws exist as torts against the harmful disclosure of private facts and constitutional protections. Early U.S. privacy law existed as a scattered tort case law following Warren and Brandeis's harms-based approach to privacy~\cite{citron2022privacy, brandeis1890right}. Privacy torts were developed in a time far before the modern technologies of the internet and mass data collection and focus on compensating for the emotional, financial, and physical harms that intrusions of privacy could cause~\cite{citron2010mainstreaming}. Cases where a recording device was secretly hidden in the bedroom to spy on intimate activities, leaking someone's social security number, or being assaulted because location data was leaked to a stalker, would give a plaintiff grounds to sue under privacy torts. Privacy torts do not cover or redress privacy intrusions conducted by database operators, harms associated with mass data collection, or issues of government surveillance. Even in instances where a plaintiff can undertake the considerable effort for suing under a privacy tort, monetary remedies, and injunctive relief are insufficient to remedy harms when data has already leaked to a broader public~\cite{iannucci2020erasing}. Additionally, courts have not always been receptive to the importance of SOGI data. LGBTQIA+ plaintiffs who file lawsuits for privacy tort offenses that involve their SOGI data often fail, even with strong cases~\cite{allen2010privacy}.

The U.S. Federal Government is limited by privacy rights granted to U.S. citizens, while agency regulators have been taking a bigger hand in promoting data privacy. The Fourth Amendment protects individuals against warrantless government surveillance that violates a reasonable expectation of privacy, which has been extended to privacy interests held by third parties~\cite{jones2020american}\footnote{\textit{Citing Carpenter v. United States}, 138 S. Ct. 2206 (2018).}. While there is no explicit right to privacy enumerated within the Constitution, the Supreme Court has found there to be a penumbra emanating from the First, Third, Fourth, Fifth, and Ninth Amendments to create a zone of privacy free from government intrusion, which has been used to justify the constitutional protection of contraceptives, same-sex intimacy, and abortion prior to the \textit{Dobbs} decision\footnote{\textit{Griswold v. Connecticut}, 381 U.S. 479 (1965)} \cite{brescia2020social}.
Neither privacy tort law nor Constitutional law offers protections against harms created by the mass collection of data or processing of it. This space has been covered by limited data protections that exist in the form of agency regulations that promise to improve protections against unfair and discriminatory practices that could infringe on the rights of queer people. The Federal Trade Commission (FTC) has recently stepped in to use its power against unfair business practices to enforce better cybersecurity and data privacy practices in the U.S. ~\cite{solove2014ftc}. When combined together, these U.S. laws and regulations form a patchwork of data protections that are not comprehensive in geography or subject and inconsistent in the level of protection and application. While both U.S. and E.U. laws promise certain rights for individuals, Waldman finds that the structure of the GDPR and other data protection schemes have been criticized for being performative, where consent is used as a shield for data extraction companies to shield themselves from accountability~\cite{waldman2022privacy}. The presence of laws and regulation is meaningless without enforcement, and privacy harms against queer people have continued unabated in the online space. 

Many of these laws and regulations will categorize many forms of biometric information as sensitive information that requires greater protection, but there is no comprehensive federal data protection scheme in the U.S. that will protect biometrics as used by big tech companies~\cite{buresh2021should}. So far, the only consumer-centric data focusing on biometric privacy is Illinois's Biometric Information Privacy Act (BIPA). Other states, like Maryland, have introduced bills modeled on the BIPA or include biometric information in broader privacy regulation~\cite{Nahra_Kane_Jessani_2023}. This leaves a gap in collecting and processing sensitive biometric information. Some laws are also limited in their regulation of gait as biometric information, such as Texas's Capture or Use of Biometric Identifier Act, whose definition of ``Biometric Identifier" does not include gait or accelerometer data. This can pose a challenge as technology advances, and the power for gait and other motion-based information becomes stronger for identification and user analytics~\cite{buresh2021should}.



\subsection{Considerations for the Queer Community}



Insufficient data protection in an age with exhaustive data extraction combines to form substantial infringements on the safety and identities of LGBTQIA+ people.  The gathering of biometric data like gait data and its subsequent power for user identification and gender recognition can expose the information of users. Information that users might not want third parties, tech companies, advertisers, data brokers, or governments to be aware of. By allowing these parties access to gender or other assumed SOGI data, other sensitive details can be presumed. Concealed gender history might be exposed to employers, coworkers, landlords, social groups, and other malicious groups who might discriminate or harass a user over their queer identity, either intentionally or subconsciously. The reconstruction of intimate personal data not only exposes sensitive information that exposes a user to harm but can also affect the determination of one's identity as it exists online. 

Social identity and the autonomy of identity are at risk when there is a constant invasion of personal privacy \cite{brescia2020social}. As Brescia argues, the integrity of identity is threatened by digital technology and platforms that seek to exploit it~\cite{brescia2020social}. This autonomy can be particularly important for queer users who wish to choose how to be identified or who to disclose to~\cite{correia2021gender,blackwell2016lgbt}. Computer vision analysis performed on a physiognomic or phrenologic basis can upset that self-autonomy and out queer people. \cite{stark2021physiognomic, katyal2021gender}

Data brokers and algorithms that only distinguish between a binary biological origin cannot accurately represent the identity of queer users. Not in how queer users interact on the platform with how they might sort them or by choosing the content to provide them. Social media platforms may undesirably provide a female-to-male transgender user with content from women-exclusive social media groups or advertisements for dresses that no longer align with their gender identity.  

The data that is collected and used to recreate identifiers of gender and queerness play into algorithmic harms and biases.  These harms have been observed in areas like algorithmic content moderation and demonetization \cite{richards2008intellectual}, the promotion and social amplification of hateful content \cite{Zitser_2021}, the normalization of hateful content, and undesirable or inaccurate outing~\cite{anderson2017glaad, ashman2020outed}.

Guzman identified how the public disclosure of SOGI data of LGBTQIA+ individuals, better known as being ``outed" in the conventional sense, can lead to discrimination, harassment, physical violence, and has historically led to the denial of rights and opportunities in the past~\cite{guzman1995outing}.
The denial of rights has long historical roots for LGBTQIA+ people.  Same-sex marriages could be banned by states, with the same laws banning gay marriage limiting marriage with transgender people by recognizing the transgender individual as their birth gender. There was no fundamental right to marriage for same-sex couples until 2015 with \textit{Obergefell v. Hodges}\footnote{\textit{Obergefell v. Hodges}, 576 U.S. 644 (2015)} \cite{larry2018transgender}. Discrimination against trans, nonbinary, and other LGBTQIA+ people is also still very present. Transgender people commonly report being harassed, physically assaulted, and economic discrimination in the workplace \cite{factor2008exploring,hill2005development}. There has been a wave of bills targeting transgender rights in the access to gender-affirming healthcare, the right to cross-dress or perform in drag, use bathrooms and other gender-segregated spaces, education about gender identity issues, updating ID cards, and other civil rights such as by removing anti-discrimination provisions \cite{Kindy_2023,Narea_Cineas_2023}. All areas could be alleviated with better data protection and privacy laws that help guarantee identity self-determination, thereby reducing the risks of privacy harms in the absence of more thorough equal protection laws.

\subsection{Gender Equal Protection, Missing Coverage}



The current federal U.S. laws protecting LGBTQIA+ people from homophobia and transphobia are limited. They are narrow to certain government-controllable sectors and often end when they might infringe on another's rights. Title VII of the Civil Rights Act of 1964  prohibited discrimination based on protected categories, including sex, but this protection did not originally extend beyond biological sex to sexual orientation or gender identity. Courts narrowly interpreted the term “sex” under Title VII to only mean a person’s sex assigned at birth. After \textit{Price Waterhouse} and \textit{Oncale}, Courts would interpret Title VII to protect transgender people under legal theories that employers discriminating against gender-nonconforming behavior would discriminate against sex ``because the discrimination would not occur but for the victim’s sex.” \footnote{\textit{Smith v. City of Salem}, Ohio, 378 F.3d 566, 574 (6th Cir. 2004).}
Over 20 states prohibit discrimination based on transgender status and sexual orientation in some form through statute. Common categories of areas where discrimination is prohibited include employment, housing, credit, public accommodations, and education \cite{NCLR_2020, mostaghim2021constructing}.

In 2020, the Supreme Court held that employment discrimination based on SOGI violated Title VII of the 1964 Civil Rights Act in \textit{Bostock v. Clayton County}. “An employer who intentionally treats a person worse because of sex—such as by firing the person for actions or attributes it would tolerate in an individual of another sex—discriminates against that person in violation of Title VII.” \footnote{\textit{Bostock v. Clayton County}, Georgia, 140 S. Ct. 1731, 1740 (2020)} However, the Supreme Court case did not decide whether First Amendment protections or religious freedom protections under the Religious Freedom Restoration Act (RFRA) which may provide exemptions to Title VII for discrimination. The Supreme Court also ruled that state public accommodation antidiscrimination law could not compel speech by businesses that disagree with LGBTQIA+ clients,\footnote{\textit{303 Creative LLC v. Elenis}, 143 S. Ct. 2298 (2023)} leaving further gaps and uncertainty in transgender protections. Nor is the protection comprehensive, and transgender claimants will still face the difficulty of proving discrimination because of their sexual orientation and gender identity. 

Because of these gaps in coverage, data privacy laws will still have significant importance in protecting the privacy of LGBTQIA+ people at the barest level, helping to prevent insidious discrimination by preventing the intrusion of private life.

\subsection{Fears and Harms. Privacy, Outing, and Digitized Dysphoria}

Researchers warned that accelerometer data infringes on user privacy because it can collect information like location and identity~\citep{kroger2019privacy}. Privacy is a tremendous concern on its own, yet using accelerometer data to predict gender identity associated with one's gait may be particularly distressing to some communities of marginalized users. Although gait is considered biometric
data revealing personal information about the owner, the federal government does not have comprehensive laws to protect the biometric data of US citizens~\citep{logan2019sale}. This has led to fears of the aforementioned privacy harms from the disclosure of SOGI data, like through public outings, but more harms derived from the use of that data: Fears of algorithmic bias, and a loss of identity self-determination. 

As of the time of this paper, there is a lack of comprehensive data privacy protections and algorithmic regulations in the U.S. to help prevent the gender recognition harms discussed in this paper. Most directly, the exposure of data is linked to privacy harms, but the inferences that can be made from raw data like gait can make inferences that become more sensitive, like SOGI data, and further inferences built on that which starts making decisions about people. What social groups do they belong to, whether they will fit in employment, the status of healthcare access, or whether their business will be refused for their identity? These are all fears that existing data privacy and equal protection laws seek to prevent for LGBTQIA+ people~\cite{FPF2022SOGI, lemberg2017hackers}. While there is a patchwork of state laws, federal regulations, and equal protection promises, there are significant concerns about growing algorithmic bias as more sophisticated and less explainable AI becomes commonplace. Harms for housing discrimination, employment discrimination, healthcare, and economic access might all happen without the direct involvement and conscious human decision against queer people, but rather the result of algorithms not designed for gender nonconformity. The insidious discriminatory effects of algorithmic bias have already been seen when comparing results between men and women, and of race \cite{lambrecht2019algorithmic, vlasceanu2022propagation, kleinberg2018discrimination}.

The lack of standards for fairness, black box design, contextual specificity, and sheer efficiency of algorithms could create the potential for ubiquitous, insidious biases for people at the margins of the U.S. population~\cite{panch2019artificial,lee2019algorithmic}. Ethnic and gender minorities are the most likely to be subject to biases, as they fail to be accurately recognized by the algorithms whose design is sampled from a less diverse set of training data~\cite{kordzadeh2022algorithmic,xiang2020reconciling}.

Algorithmic biases and AI discrimination are areas that the US government has signaled intent on regulating, potentially through interpretations of anti-discrimination law as well as recent informal guidance. The FTC has released informal guidance that they intend to use their powers to target AI discrimination as illegal, unfair, deceptive acts and practices~\cite{jillson2021aiming}. A combined board with representatives of the FTC, Consumer Finance Protection Bureau, Equal Employment Opportunity Division, and the Justice Department's Civil Rights Division have made statements of their intent to enforce civil rights and non-discrimination, and other legal protections \cite{FederalTradeCommission_2023}. 

The effects of algorithmic bias on queer people are understudied. However, the direct, personal effects that gender recognition and other algorithms can have are more present. The labels and classification results of automated gender recognition have the potential to directly misgender trans and queer users, and induce third parties to introduce dysphoric materials like undesirable advertising~\cite{keyes2018misgendering, hamidi2018gender}. AI and automated decisionmaking systems that adhere to a cisnormative design pose the risk and harms of digitizing dysphoria as a constant in queer users' lives. 

While a few~\citep{singh2019side,kroger2019privacy} have brought attention to privacy concerns, there are issues with applying biometrics to non-binary personages that create the loss of nuanced identity data~\citep{clarke2018they,albert2021whole}. Therefore, smartphone sensor data for gender prediction risks overgeneralizing gender norms, exacerbating privacy concerns, and inducing dysphoric implications on queer bodies. 

Throughout the decades, most of the research around biometric gait recognition has primarily focused on accuracy and efficiency~\cite{gafurov2007survey}, and this largely remains true today. While tremendous advances and strides have been made to improve gait recognition with the progress of machine learning and artificial intelligence, there is more emphasis on the new advancements in technology and absence or glossing over whether we should do this type of work given the technological harms and reinforcing gendered stereotypes.

Revisiting Mantyjarvi~\cite{1415569} from earlier, they found promising results in unlocking the phone by assessing movement, such as gait, by accessing phone accelerometers. However, the researchers found difficulties and potential drawbacks due to how one's gait could be changed based on footwear and ground surfaces, as Mantyjarvi \textit{et al.} have noted in their study that gait changes based on footwear, but further nuance that footwear can be fluid~\cite{1415569}. Gafurov has also pointed out that the security of biometric gait recognition has not been studied~\cite{gafurov2007survey}. Bouchrika later argued that using gait would be "a more suited modality for people recognition in surveillance and forensic scenarios...due to non-intrusively and covertly from a distance even with poor resolution imageries."~\cite{bouchrika2018survey}. These all reveal potential failures and inaccuracies in the creation and interpretation of gait data, leading to inaccuracies in gender recognition algorithms. The privacy harms, algorithmic biases, and digitized dysphoria can be caused by unfettered and inaccurate predictive algorithms based on gait recognition. 

\section{Food for Thought \& Queer Futures}
\label{sec:foodforthought}
\subsection{Codifying Inclusive Law}
Some proposals promise to regulate algorithms and AI, focusing on algorithmic bias and discrimination, but most are only in their early stages and have not taken effect. The E.U. AI Act has been the furthest reaching. It is structured by differentiating between purposes. AI systems with low-risk purposes, like spam filters, require limited disclosure. Those of high risk, such as AI for autonomous vehicles and medical devices, demand greater testing, transparency, and accountability. The E.U. AI Act lastly finds that some purposes, like using AI for a social credit scoring system like in China, are deemed unacceptable under the AI Act \cite{TheArtificialIntelligenceAct_2023, Feingold_2023}. Yet, these beginning proposals are still missing the inclusiveness of queer communities in the international law landscape. 

The U.S. has seen several proposals, with the Biden Administration releasing a Blueprint for an AI Bill of Rights that covers algorithmic discrimination and data privacy protections while incorporating other concepts for data protection~\cite{TheWhiteHouse_2023}.
Legislative action taken to curb AI harms has been proposed in the federal government, with the Algorithmic Accountability Act of 2019 being introduced by Senator Ron Wyden and updated in 2022. While it features protections against algorithmic discrimination and greater requirements for impact assessment, neither bill has proceeded far in the legislative process~\cite{SenatorRonWydenofOregon_2022, HouseBill6580}.

Regulatory agencies do have a prominent role at least, with the FTC taking a strong position to regulate AI creation and algorithmic use under its focus on unfair business practices, as well as credit reporting and equal protection powers. A recent FTC settlement has required the deletion of a facial recognition algorithm developed from user data without consent  \cite{Vedova_Technology_Jillson_Fair_2022, Gesser_Rubin_Gesser_2022}.
 This punitive measure has the potential to be a strong deterrent to bad data practices, but it has yet to be seen if the FTC will begin regulating biased algorithms in the same way. 

The creation of these new bills and regulations would hopefully redress the queer community's growing concerns of algorithmic bias and unrestrained AI development and use that is not designed for LGBTQIA+ people. Algorithmic and privacy harms currently theorized could potentially escalate and rise in frequency, creating an online ecosystem where queer people face automated discrimination and digitized dysphoria when interacting with apps and websites whose algorithms are only designed for a gender binary. Regulation on how data can be captured and algorithms can be created to be more equitable and acknowledge gender minorities. While new regulations and laws in the U.S. might be inevitable, technology developers and industry do not need to rely on or wait for their creation before becoming more inclusive.

\subsection{Building Inclusive Technologies}

Our paper highlights significant gaps in gender inclusivity within biometrics and broader AI-driven technology. For instance, choosing to design a gait recognition task based on mobile device placement on one's body is one of many examples which not only infringe upon user privacy but also perpetuate gender normative design. Choosing to control for clothing in gait recognition is another myopic example by which binary gender assumptions propagate into the erasure of non-hegenomic identities. In addressing these gaps, an opportunity presents itself: ML researchers and technopolicy makers alike can move towards addressing privacy and representational harms through critical discourse surrounding the social norms driving their respective sociotechnical design choices. In our provided examples, this looks like taking a moment to question biometric approaches infused with cis and heteronormativity and contextualizing their impact on gender minorities.

Several scholars provide avenues by which wearable technology can intentionally support the existence of queer bodies \citep[\textit{inter~alia}]{albert2021whole, bolesnikov2023wearable}. These works, among many others, serve as examples for reimagining inclusive biometric technologies. Furthermore, being that ML researchers and technopolicy stakeholders co-exist within the broader AI ecosystem, we highly encourage open and honest conversation on ways in which AI harms can directly result from a lack of diversity in inclusion criterias which shape AI-driven systems. Asking these questions may be facilitated through the use of reflexivity, a vehicle to investigate how one's own biases, values, and social locations are imparted onto a given process. Here,  taking into account one's positionality can help illuminate gaps in coverage during the design of task definition in data-driven processes and legislature alike \cite{boyd2021quantitative, Boyd2023reflexive, ovalle2023factoring}. 

Dr. Ian Malcolm's quote highlights a tension we have explored throughout this paper which grapples with access to large-scale mobile biometrics, the power of choice, and the consequent possibilities to do harm. While mobile biometrics can add value to daily life, there are many contexts in which operationalizing biometric data erases non-hegemonic identities, evidenced earlier by the ways gender prediction both reinforces stigmas about queer bodies while erasing them. Dr. Malcolm's quote reminds us to deliberately scrutinize assumptions behind decisions influencing AI ecosystems as whole. Even if we can leverage mobile biometrics to infer a ``gender label'' via  clothes segmentation techniques and smartphone placement on the body  --- \textit{should we?}

\bibliographystyle{ACM-Reference-Format}


\bibliography{references}

\appendix

\end{document}